\documentclass[twocolumn]{aastex62}
\PassOptionsToPackage{table}{xcolor}
\usepackage[table]{xcolor}
\usepackage{tabularx,colortbl}
\usepackage{graphicx}
\usepackage{color,colortbl}
\usepackage{amsmath}
\usepackage{lineno}
\usepackage{multirow}
\usepackage{enumitem}
\usepackage{graphicx}
\usepackage{dcolumn}
\usepackage{bm}
\usepackage{hyperref}
\usepackage{enumitem}

\definecolor{bittersweet}{rgb}{1.0, 0.44, 0.37}

\begin{document}

\title{{\bf Micro Tidal Disruption Events in Active Galactic Nuclei}}

\author{Y. Yang}
\affiliation{Department of Physics, University of Florida, PO Box 118440, Gainesville, FL 32611-8440, USA}

\author{I. Bartos}
\thanks{imrebartos@ufl.edu}
\affiliation{Department of Physics, University of Florida, PO Box 118440, Gainesville, FL 32611-8440, USA}

\author{G. Fragione}
\affil{ Department of Physics \& Astronomy, Northwestern University, Evanston, IL 60208, USA}
\affil{Center for Interdisciplinary Exploration \& Research in Astrophysics (CIERA), Northwestern University, Evanston, IL 60208, USA}

\author{Z. Haiman}
\affiliation{Department of Astronomy, Columbia University, 550 W. 120th St., New York, NY, 10027, USA}

\author{M. Kowalski}
\affiliation{Deutsches Elektronen Synchrotron DESY, Platanenallee 6, 15738 Zeuthen, Germany}
\affiliation{Institut fur Physik, Humboldt-Universit\:at zu Berlin, D-12489 Berlin, Germany}

\author{S. M\'arka}
\affiliation{Columbia Astrophysics Laboratory, Columbia University in the City of New York, New York, NY 10027, USA}

\author{R. Perna}
\affiliation{Department of Physics and Astronomy, Stony Brook University, Stony Brook, NY 11794-3800, USA}
\affiliation{Center for Computational Astrophysics, Flatiron Institute, New York, NY 10010, USA}

\author{H. Tagawa}
\affiliation{Astronomical Institute, Graduate School of Science, Tohoku University, Aoba, Sendai 980-8578, Japan}

\begin{abstract}
Active galactic nuclei (AGNs) can act as black hole assembly lines, funneling some of the stellar-mass black holes from the vicinity of the galactic center into the inner plane of the AGN disk where the black holes can merge through dynamical friction and gravitational wave emission. Here, we show that stars near the galactic center are also brought into the AGN disk, where they can be tidally disrupted by the stellar-mass black holes in the disk. Such {\it micro-tidal disruption events} (micro-TDEs) could be useful probe of stellar interaction with the AGN disk. We find that micro-TDEs in AGNs occur at a rate of $\sim170$\,Gpc$^{-3}$yr$^{-1}$. Their cleanest observational probe may be the detection of tidal disruption in AGNs by heavy supermassive black holes ($M_{\bullet}\gtrsim10^{8}$\,M$_{\odot}$) so that cannot tidally disrupt solar-type stars. We discuss two such TDE candidates observed to date (ASASSN-15lh and ZTF19aailpwl).
\vspace{1cm}
\end{abstract}

\section{Introduction}

A few percent of galaxies host a central, compact, highly luminous region called active galactic nucleus (AGN). AGNs are the result of a highly accreting supermassive black hole (SMBH), in which infalling gas forms an accretion disk around the SMBH. 

In addition to being the source of the high central luminosity, AGN disks can also impact the dynamics of stellar remnants in the galactic center. Of particular interest has been the interaction of AGN disks with stellar-mass black holes (BHs) that have migrated to the galactic center through mass segregation \citep{2009ApJ...697.1861A,2009ApJ...698L..64K,2009MNRAS.395.2127O}. While orbiting the central SMBH, BHs periodically cross the AGN disk that gradually aligns their orbit with the disk plane \citep{Bartos17}. Once in the disk, BHs migrate inwards due to 
viscous-tidal interactions with the disk \citep{McKernan12,Bellovary16}. After these processes brought two BHs in each others' vicinity, the dense gas of the AGN disk facilitates gravitational capture and ultimately the binary merger of the two BHs through dynamical friction. 

BH mergers in AGN disks may be an important gravitational-wave source with BH properties distinct from those expected from other astrophysical mechanisms \citep{Bartos17,2017MNRAS.464..946S,2018ApJ...866...66M,Yang19a,Yang19b_PRL,Yang20,Yang20_GW190814,2019ApJ...878...85S,Tagawa19,2020ApJ...899...26T,Tagawa20_ecc,Tagawa20_massgap,Gayathri20_GW190521,Samsing20}. Merging BHs or neutron stars in AGN disks might also produce detectable electromagnetic emission, opening up another way to study AGN-assisted mergers \citep{2019ApJ...884L..50M, 2021arXiv210302461K, Perna2021,2021arXiv210310963P,2021arXiv210409389Z,2021ApJ...906L..11Z}.

Stellar orbits are also affected by AGN disks as disk crossings dissipate orbital energy~\citep{2000ApJ...545..847M}, which will be particularly significant for stars on highly eccentric orbits. Some of these stars will gradually get closer to the central SMBH until they are tidally disrupted by it. Nonetheless such interactions with the AGN disk may have limited effect on the overall tidal disruption rate in galaxies with AGNs  (\citealt{2020ApJ...889...94M}, but see \citealt{Tadhunter2017}). 

Here we examine the orbital alignment of stars with the AGN disk, and its consequences. Similarly to BHs and neutron stars, some of the stars in galactic centers will align their orbits with the AGN disk plane. Once in the plane, stars can form binaries with BHs, leading to their eventual tidal disruption within the AGN disk. We discuss the rate density of such tidal disruption events around BHs (hereafter micro-TDEs; \citealt{2016ApJ...823..113P}), their expected observational signatures and possible observational indications of their existence. 

\section{Orbital evolution around AGN disks}

Upon crossing the AGN disk, the velocity of stars will change due to the accretion of matter from the disk. We simulated the orbital evolution of stars accounting for this velocity change following \cite{2019ApJ...876..122Y}. The mass of infalling gas upon each crossing is given by
\begin{equation}
    \Delta m_{\rm gas}=v_{\rm rel}t_{\rm cross}R_{\rm cap}^2\pi\Sigma/(2H)\,,
\label{eq:mcross}
\end{equation} 
where $R_{*}$ and $M_{*}$ are the stellar radius and mass, 
$R_{\rm cap}=\max\{R_{*}, R_{\rm BHL}\}$, 
$R_{\rm BHL}\equiv 2GM_{*}/(v^2_{\rm rel}+c_s^2)$ is the stars' Bondi-Hoyle-Lyttleton radius, $\Sigma$ and $H$ are the surface density and scale height of the AGN disk respectively, $t_{\rm cross}\equiv 2H/v_{\rm z}$ is the crossing time, $v_{\rm z}$ is the $z$ component of the stars' velocity. We found mass loss by 
a main-sequence star due to ram pressure exerted by the disk gas to be negligible \citep{2005ApJ...619...30M}.

We considered a geometrically thin, optically thick, radiatively efficient, steady-state
\citep{1973A&A....24..337S} but self-gravitating \citep{2003MNRAS.341..501S} accretion disk. We adopted viscosity parameter $\alpha=0.1$, accretion rate $\dot{M}_{\bullet}=0.1\dot{M}_{\rm Edd}$ and radiation efficiency $\epsilon=0.1$. We used a fiducial SMBH mass of $10^6$\,M$_{\odot}$.

We computed the changes in the velocity and angular momentum of the star upon each crossing based on momentum conservation \citep{2019ApJ...876..122Y}:
\begin{align}
\bf{\Delta {v_{*}}}&=-\lambda(\bf{v_{*}}-\bf{v_{gas}})=-\lambda \bf{v_{\rm vel}}\\
\bf{\Delta J}&=-\lambda \bf{r}\times \bf{v_{\rm rel}}\,, 
\end{align}
where $\bf{v_{\rm rel}} $ is the relative velocity between the star and the gas, and $\lambda\equiv\Delta M_{\rm gas}/M_{*}$ is a dimensionless factor. We updated the parameters of the star's orbit after each crossing according to the change of velocity and obtained the orbital evolution iteratively.

We assumed that the stars follow a mass-radius relation $R_{*}=1.06(M_*/{\rm M}_\odot)^{0.945}$ for $M_{*}<1.66M_{\odot}$ and $1.33(M_*/{\rm M}_\odot)^{0.555}$ for $M_{*}>1.66M_{\odot}$ \citep{1991Ap&SS.181..313D}. We assumed that the stellar mass follows a Kroupa initial mass function \citep{2001MNRAS.322..231K}. We took into account the mass segregation in the spatial distributions of main sequence stars, adopting $dN/da\propto a^{-3/2-0.5M_{*}/M_{\rm max}}$, where $a$ is the semi-major axes of the star's orbit around the SMBH \citep{2009ApJ...698L..64K,2009ApJ...697.1861A,2018ApJ...860....5G}. We adopted a maximum stellar mass of  $M_{\rm max}=50$\,M$_{\odot}$, which accounts for the fact that the lifetime of the most massive stars is too short for them to participate in mass segregation. The total stellar mass within the gravitational influence radius of the SMBH was taken to be equal to the mass of the SMBH \citep{2000ApJ...545..847M}.

We similarly computed the orbital alignment of BHs and neutron stars with the AGN disk. For simplicity we adopted the Salpeter initial mass function $dN/dM\propto M^{-2.35}$ for BHs within the range of $5-50$\,M$_\odot$ and a normal initial mass function $M/M_{\odot}\sim\textit{N}(1.49,0.19)$ for neutron stars \citep{2016ARA&A..54..401O}. We assumed that the total mass of the BH population is $1.6\%$ of the stellar mass in galactic centers and the number of neutron stars is ten times the number of BHs \citep{2018ApJ...860....5G}. We additionally took into account mass segregation in the three-dimensional spatial distributions of BHs and neutron stars following \cite{2018ApJ...860....5G}. Otherwise orbital alignment was computed similarly to stars.

\section{Binary formation, evolution and tidal disruption}

Once some of the stars, BHs and neutron stars align their orbits with the AGN disk, they begin migrating inward 
analogously to planetary migration in protostellar disks (e.g., \citealt{2020ApJ...899...26T}). Both orbital alignment and migration increase the number density of each type of object in the inner AGN disk, leading to efficient binary formation through gravitational capture. Dynamical friction within the gas further facilitates the formation of binaries and their consecutive hardening.

We considered a binary consisting of a main sequence star and a BH. The binary typically forms at small separations of only $\sim 10$ times the star's tidal radius $R_{\rm t} = R_*(M_{\rm bh}/M_*)^{1/3}$, and could be eccentric at (or soon after) its birth, either due to its formation (c.f. \citealt{2004Natur.427..518F}) and/or subsequently driven to be eccentric by a circumbinary disk \citep{2021ApJ...909L..13Z,2021arXiv210309251D}. This could justify an impulsive disruption "event" at the pericenter of an ellipse, rather than a slow circular inspiral and gradual tidal "peeling".

Once the binary's pericenter distance $R_{p}$ approaches $R_{\rm t}$, the star is tidally disrupted. TDEs around SMBHs typically occur at parabolic ($e\approx1$) stellar orbits, leading to a tidal tail that will affect long-term accretion. In addition, partial disruption has a substantially larger cross-section in the case of TDEs, and should be more common \citep{2020SSRv..216...35S}. The lower eccentricity of micro-TDEs in AGNs will result in most of the stellar material remaining in the vicinity of the BH, leading to a more massive and compact accretion disk. 

To qualitatively understand the above tidal disruption process, we performed a simulation of a micro-TDE using smoothed-particle hydrodynamics (SPH;  \citealt{2010ARA&A..48..391S}). We considered a star with mass $M_*=1$\,M$_\odot$ and a BH with mass $M_{\rm bh}=10$\,M$_\odot$. The star was initially located at $5R_{\odot}$ distance from the BH.  For simplicity, we assumed that the BH-star binary is on a circular orbit.

We adopted a polytrope model to compute the structure of the main sequence star, assuming that pressure depends on the density via $P=k\rho^{1+1/n}$, where constant $k$ is chosen to produce a Sun-like star and $n$ is set to be 3, which can describe the structure of a main sequence star. We considered the self gravity and adopted artificial viscosity in our simulation \citep{2010ARA&A..48..391S}. 

\begin{figure*}
\begin{center}
	\includegraphics[scale=0.7]{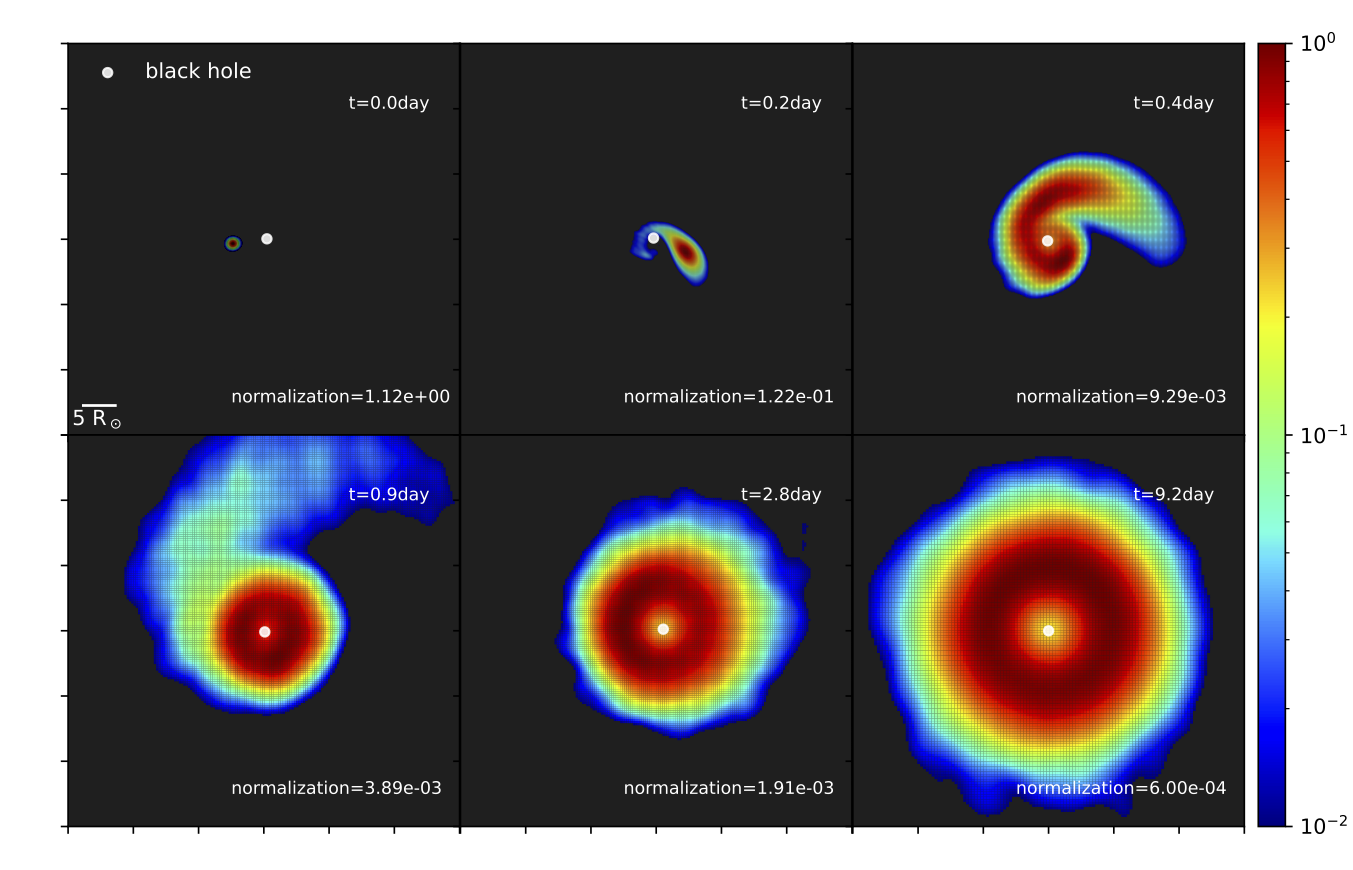}
    \caption{{\bf Smoothed-particle hydrodynamics simulation of micro-TDE of a binary system} consisting of a 10$M_{\odot}$ BH and a 1$M_{\odot}$ main sequence star, the initially circular orbit with 5$R_{\odot}$ separation. The period of the initial orbit is about 0.4\,days. We show the normalized surface density of the debris at different times and give the normalization factor in each subplot. }
    \label{fig:TDE_sph}
\end{center}
\end{figure*}

The outcome of our simulation is shown in Fig \ref{fig:TDE_sph}. We see that tidal instability rapidly grows and the star is tidally disrupted within a day. Most of the stellar matter remains gravitationally bound to the BH. 

The above simulation does not account for the presence of the mini-disk around the BH accumulated from the AGN disk prior to stellar disruption. This can be quantitatively important but we do not expect it to qualitatively change the outcome. We estimated that stellar mass loss prior to disruption due to radiation by the accreting BH will be only $\sim 10\%$ of the star's mass and therefore will not meaningfully affect the outcome \citep{2020ApJ...903..140Z,2021arXiv210302461K}.

\section{Radiation features of micro-TDEs}

TDE and micro-TDE light curves and spectra are not yet well understood. They also depend on a large number of variables, such as the masses of the black hole and the star, the eccentricity and impact parameter of the encounter, the black hole's spin, and others. 

While in some ways similar, TDE and micro-TDE light curves and spectra can differ due to multiple effects. First, TDEs by SMBHS occur with highly eccentric orbits ($e\approx1$), while AGN-assisted micro-TDEs are not expected to be highly eccentric. This latter case allows a larger fraction of the stellar matter to remain in the vicinity of the BH, quickly forming a nearly circular accretion disk \citep{2019ApJ...881...75K}. Second, the larger amount of mass and lower black hole mass in the micro-TDE case results in highly super-Eddington accretion that can exceed the Eddington rate $\dot{M}_{\rm Edd}\equiv L_{\rm Edd}/c^2=2.6\times10^{-9}M_{\rm BH}$\,yr$^{-1}$ by up to a factor of $10^5$ \citep{2019ApJ...881...75K}. Here, $L_{\rm Edd}$ is the Eddington luminosity and $c$ is the speed of light. Super-Eddington accretion results in a geometrically thick accretion disk. A large fraction of the disk mass is expected to be blown away as a disk wind. Radiation from the inner accretion disk will escape with a delay, resulting in an extended peak in the light curve \citep{2019ApJ...881...75K}. Further delay in reaching peak luminosity may be expected given that the AGN will surround micro-TDEs with a dense medium. The delay, and even whether the micro TDE emission can escape the AGN disk, may depend on the mass of the AGN disk and the location of the disruption event within the disk \citep{Perna2021}. In our fiducial disk model, micro-TDEs are expected mostly in regions around $10^{-4}-10^{-2}$\,pc from the central SMBH, where the disk is ionized and the opacity to scattering is very high, significantly altering the observed emission. However, BHs in the AGN disk are expected to open cavities due to accretion-induced radiation \citep{2021arXiv210302461K}, which results in the reduction of opacity, leaving emission from the micro-TDE less affected. Finally, the vicinity of the SMBH for binaries that reach the AGN disk's migration trap could also further alter the expected light curves.

In the case of micro-TDEs in AGNs, a large fraction of the stellar mass remains gravitationally bound to the BH, resulting in a bolometric luminosity of up to $\sim 10^{44}$\,erg\,s$^{-1}$ \citep{2019ApJ...881...75K}. By comparison, the amount of stellar matter available in the case of TDEs depends on the penetration parameter $\beta\equiv R_{\rm t}/R_{\rm p}$. Partial disruption ($\beta \lesssim 1$) is more common, resulting in lower luminosity \citep{2020SSRv..216...35S}.

The long-term evolution of TDEs can be characterized by a smooth power-law decay \citep{2021ApJ...908....4V}, with a fiducial power-law index of $-5/3$ \citep{1988Natur.333..523R}. Micro-TDEs are expected to result in similar power-law decays in the long term \citep{2019ApJ...881...75K}.
SPH simulations of micro-TDEs found the power-law index to vary from $t^{-5/3}$ to $t^{-9/4}$, with steeper indices corresponding to lower BH mass \citep{Wang2021}. 

The high accretion rate in micro TDEs might also launch relativistic outflows that produce significant $\gamma$-ray and X-ray emission. Such emission would be weaker and longer than typical GRBs, possibly resembling that of ultralong GRBs \citep{2016ApJ...823..113P}. Such $\gamma$ and X-ray emission is highly beamed and is therefore only detectable from a fraction of micro TDEs. Micro TDE-driven GRBs could be differentiated from other types of GRBs through the identification of a thermal TDE-like counterpart with longer-wavelength observations, and possibly directional coincidence with AGNs.

\section{SMBHs too heavy for TDEs}

For SMBHs whose Schwarzschild radius is greater than their tidal radius, stars will be swallowed whole before they can be tidally disrupted. The maximum (non-rotating) SMBH mass capable of producing TDEs, called the Hills mass \citep{1975Natur.254..295H}, is around $10^{8}$\,M$_\odot$ for solar-type stars \citep{2020SSRv..216...35S}. This limit is only weakly dependent on the stellar mass for zero-age main sequence stars, although it can be greater for off-main-sequence giant stars, or for highly spinning SMBHs.

Micro-TDEs in AGNs are not limited by the SMBH mass. Therefore, given the uncertainties in their light curves and spectra, the identification of TDE-candidates in AGNs with SMBH mass beyond $10^{8}$\,M$_\odot$ may help distinguish micro-TDEs and TDEs. 

There has been two identifications of TDE-like emission from AGNs with SMBH mass $M_\bullet>10^{8}$\,M$_\odot$: 

{\bf ASASSN-15lh} was an unusually bright transient first discovered by the All-Sky Automated Survey for Supernovae (ASAS-SN; \citealt{2014ApJ...788...48S}) in 2015 at a redshift of $z = 0.232$ \citep{ASASSN-15lh}. The spectrum of ASASSN-15lh points to a TDE and rules out superluminous supernova origin \citep{2016NatAs...1E...2L,2018A&A...610A..14K}. ASASSN-15lh was found to come from a galactic nucleus with SMBH mass $M_\bullet=5^{+8}_{-3}\times10^8$\,M$_\odot$ \citep{2018A&A...610A..14K}, which may be a low-luminosity AGN based on its BPT classification \citep{2018A&A...610A..14K}. This is beyond the Hills mass for solar-type stars, although a highly spinning SMBH in this mass range could produce a TDE. The late-time ($\gtrsim 100$\,days) light curve and spectrum of ASASSN-15lh could be well explained with a close to maximally spinning SMBH with mass $M_\bullet\sim10^9$\,M$_\odot$, although the early light curve and the lack of radio and dim X-ray emission are not understood \citep{2020MNRAS.497L..13M}. The peak luminosity of ASASSN-15lh was $L_{\rm peak}\sim 5\times10^{45}$\,erg\,s$^{-1}$, higher than the $\sim 10^{44}$\,erg\,s$^{-1}$ predicted from analytical micro-TDE models \citep{2016NatAs...1E...2L}, which nevertheless have large uncertainties.

{\bf ZTF19aailpwl} was a bright ($L_{\rm peak}\sim 10^{45}$\,erg\,s$^{-1}$) transient first observed by the Zwicky Transient Facility \citep{2019PASP..131g8001G} in 2019 at a redshift of $z = 0.37362$ \citep{2020arXiv201008554F}. It originated from an AGN with a reconstructed SMBH mass of $M_\bullet\sim10^{8.2}$\,M$_\odot$ (although a mass $<10^{8}$\,M$_\odot$ cannot be completely ruled out;  \citealt{2019PASP..131g8001G}). Follow-up observations found that the spectral properties of ZTF19aailpwl are consistent with a TDE \citep{2019PASP..131g8001G}, but also with an AGN flare attributed to enhanced accretion reported by \cite{2019NatAs...3..242T}. Its light curve is similar to that of TDEs, with a somewhat longer than usual rise time. For micro-TDEs, such longer rise time may be possible due to interaction with the disk wind and the low eccentricity of the stellar orbit that makes the encounter less impulsive and the tidal effects more gradual, compared to the usual SMBH-TDE case.

For both cases above, the reconstructed black hole masses beyond the Hills mass point away from TDEs but are consistent with a micro-TDE origin. 

Micro-TDEs can also occur outside of AGN disks, for example in stellar triples hosting a BH \citep{2019MNRAS.489..727F}, or in dense stellar clusters \citep{2019ApJ...881...75K,2019PhRvD.100d3009S,2021MNRAS.500.4307F}. The rate of micro-TDEs from these channels is uncertain but could be as high as that of TDEs. These micro-TDEs could also occur in galaxies with SMBH mass above the Hills mass. Observationally, however, these micro-TDE channels will occur far from galactic centers and will not be confined to AGNs. As both ASASSN-15lh and ZTF19aailpwl were localized to the galactic center and in an AGN means that, for these two cases, non-AGN-assisted micro-TDE channels are unlikely.

\section{Rate Density}

To estimate the expected rate density of micro-TDEs, we computed the number of stars, BHs and neutron stars that undergo orbital alignment with the AGN disk within the AGN's lifetime (assumed to be $10^7$\,yr). We found $\sim 5$ orbital alignments for each object type (see also \citealt{2020ApJ...901L..34Y}). We simulated the merger of objects within the disk using a Monte Carlo simulation of many galaxies, with random number of orbital alignments following a Poisson distribution for each galaxy. 

To compute the rate density we further assumed for simplicity that the expected orbital alignment rate is identical in every AGN (there is only a weak dependence on the AGN properties; \citealt{2019ApJ...876..122Y}). We adopted a number density $n_{\rm seyfert}=0.018$\,Mpc$^{-3}$ for Seyfert galaxies \citep{2005AJ....129.1795H} which represent most AGNs. We show the resulting expected merger rate density for all possible combinations of stars, BHs and neutron stars in Table \ref{table :merger rate}.

\begin{table}[!htbp]
\begin{center}
\begin{tabular}{c|c|c|c|}
\cline{2-4}
& black hole & neutron star & star  \\
\hline
\multicolumn{1}{|c|}{black hole} & 13 & 1 & 170 \\
\hline
\multicolumn{1}{|c|}{neutron star} & & 10$^{-3}$ & 0.14 \\   
\hline
\multicolumn{1}{|c|}{star} & & & 20 \\
\hline
\end{tabular}
\caption{{\bf Expected rate density of binary encounters in AGNs}. Micro-TDEs correspond to BH--star encounters. Results are shown in units of Gpc$^{-3}$ yr$^{-1}$. While BHs are the least common, they dominate the merger rate as their higher mass results in more efficient mass segregation and orbital alignment, while their mergers result in BHs that can undergo further mergers in the AGN disk.}
\label{table :merger rate}
\end{center}
\end{table}

We separately estimated the micro-TDE rate density from observations. For this we used the rate density of TDE candidates in galaxies with $M_\bullet\gtrsim10^{8}$\,M$_\odot$, which are difficult to explain with disruption by SMBHs. We used the TDE host galaxy SMBH mass function estimated by \cite{2018ApJ...852...72V} based on about two dozen observed TDE-candidates. \cite{2018ApJ...852...72V} found that the mass function is roughly constant for $M_\bullet < 10^{7.5}$\,M$_\odot$, while it sharply drops for $M_\bullet > 10^{7.5}$\,M$_\odot$ (in this high mass range they only consider the detection of ASASSN-15lh). Their reconstructed mass function corresponds to a TDE rate of $5^{+8}_{-4}\times10^{-2}$\,Gpc$^{-3}$yr$^{-1}$ (1$\sigma$ uncertainty) for $M_\bullet > 10^{8}$\,M$_\odot$. 
Taking into account the mass function of SMBHs in AGNs, we found that about $1\%$ of micro-TDEs occur in AGNs with $M_\bullet > 10^{8}$\,M$_\odot$, corresponding to a rate of $\sim 2$\,Gpc$^{-3}$yr$^{-1}$. 


\section{Conclusion}

We proposed that micro-TDEs occur in AGN disks, and found their rate to be about $170$\,Gpc$^{-3}$yr$^{-1}$. Such micro-TDEs may be easiest to distinguish from TDEs around SMBHs by focusing on AGN-hosting galaxies in which the central SMBH's mass is too high ($M_\bullet\gtrsim10^{8}$M$_\odot$) to tidally disrupt solar-like stars. Two such TDE candidates have been reported so far, ASASSN-15lh and ZTF19aailpwl, both among the highest-luminosity TDE candidates. 

The observed rate density of TDE candidates from galaxies with $M_\bullet\gtrsim10^{8}$M$_\odot$ \citep{2018ApJ...852...72V} is below the micro-TDE rate density of $\sim2$\,Gpc$^{-3}$yr$^{-1}$ predicted in this work for AGNs with $M_\bullet\gtrsim10^{8}$M$_\odot$, and might be a bright sub-population. In addition, one candidate (ZTF19aailpwl) occurred in an AGN, while the other (ASASSN-15lh) probably occurred in a weak AGN. 

The unique environments in AGN disks are expected to give rise to a wealth of other interesting phenomena, such as the tidal disruption of a star by a neutron star (albeit at a smaller rate of $0.1$\,Gpc$^{-3}$yr$^{-1}$), the formation of Thorne-Zytkow objects \citep{1975ApJ...199L..19T} when a neutron star in the disk is unable to tidally disrupt a stellar companion due to a large stellar radius, the tidal disruption of white dwarfs by stellar-mass BHs, with unique signatures~\citep{Maguire+2020}, and the accretion induced collapse of neutron stars \citep{2021arXiv210310963P} and white dwarfs \citep{2021arXiv210409389Z}.

Further theoretical and observational work is needed to better understand the spectral and temporal properties of AGN-assisted micro-TDEs, and to observe them against the background of AGN variability. In particular we suggest an observational focus on AGNs that harbor the heaviest SMBHs and exhibit unusual flaring activity (e.g., \citealt{2019NatAs...3..242T}).

\begin{acknowledgments}
The authors would like to thank 
Sjoert van Velzen for useful suggestions. I.B. acknowledges the support of the the Alfred P. Sloan Foundation. ZH was supported by NASA grant NNX15AB19G and NSF grants AST-2006176 and AST-1715661. HT acknowledges support by the Grants-in-Aid for Basic Research by the Ministry of Education, Science and Culture of Japan (HT:17H01102, 17H06360). RP acknowledges support by NSF award AST-2006839 and from NASA (Fermi) award 80NSSC20K1570.
\end{acknowledgments}

\bibliography{reference}
\end{document}